\documentclass{PoS}
\usepackage{amssymb,amsmath}
\usepackage[labelfont=bf,textfont=it]{caption}
\allowdisplaybreaks

\title{Results from TopFitter}

\ShortTitle{Results from TopFitter}

\author{Andy Buckley\\
       SUPA, School of Physics and Astronomy, University of Glasgow, Glasgow, G12 8QQ, United Kingdom\\
        E-mail: \email{andy.buckley@glasgow.ac.uk}}

\author{Christoph Englert\\
       SUPA, School of Physics and Astronomy, University of Glasgow, Glasgow, G12 8QQ, United Kingdom\\
       E-mail: \email{christoph.englert@glasgow.ac.uk}} 

\author{James Ferrando\\
  DESY, D-22607 Hamburg, Germany\\
  E-mail: \email{james.ferrando@desy.de}} 
        
\author{David J. Miller\\
  SUPA, School of Physics and Astronomy, University of Glasgow, Glasgow, G12 8QQ, United Kingdom\\
  E-mail: \email{david.j.miller@glasgow.ac.uk}}        

\author{Liam Moore\\
  Centre for Cosmology, Particle Physics and Phenomenology (CP3), Universit\'e catholique de Louvain,
  B-1348 Louvain-la-Neuve, Belgium\\
  E-mail: \email{l.moore@cern.ch}}    
  
\author{Karl Nordstr\"om\\
  SUPA, School of Physics and Astronomy, University of Glasgow, Glasgow, G12 8QQ, United Kingdom\\
  E-mail: \email{k.nordstrom.1@research.gla.ac.uk}}       
        
\author{\speaker{Michael Russell}\\%
  SUPA, School of Physics and Astronomy, University of Glasgow, Glasgow, G12 8QQ, United Kingdom\\
  E-mail: \email{m.russell.2@research.gla.ac.uk}}

\author{Chris D. White\\
      Centre for Research in String Theory, School of Physics and Astronomy,
      Queen Mary University of London, 327 Mile End Road, London E1 4NS, UK\\
      E-mail: \email{christopher.white@qmul.ac.uk}}    
        
\abstract{We discuss a global fit of top quark BSM couplings, phrased in the model-independent language of higher-dimensional effective operators, to the currently available data from the LHC and Tevatron. We examine the interplay between inclusive and differential measurements, and the complementarity of LHC and Tevatron results. We conclude with a discussion of projections for improvement over LHC Run II.}

\FullConference{ 9th International Workshop on the CKM Unitarity Triangle\\
		 28  November - 3 December 2016\\
		 Tata Institute for Fundamental Research (TIFR), Mumbai, India}

\begin{document}

\section{Introduction}
The lack of convincing signals of new physics in the data from the Tevatron and LHC has led to the resurgence of model-independent frameworks for parameterising deviations from Standard Model predictions. It also suggests that if there is new physics, there is either a large separation in energy between the scale that it resides at; $\Lambda$, and the scales probed by the LHC; $E$, or that it is very weakly coupled, or both. In either case, the Standard Model can be viewed as the leading set terms in an effective field theory (EFT), where the non-SM interactions are captured by an infinite series of higher-dimensional (that is, $D > 4$) operators, which, by virtue of their renormalisation group scaling, become increasingly important at higher energy scales. The leading effects on collider observables enter at dimension $D=6$, and the series is typically truncated at the same order, which is justified provided that $ E \ll \Lambda$. The Standard Model EFT can then be succinctly summarised as

\begin{equation}
\mathcal{L}_{\text{full}} = \mathcal{L}_{\text{SM}} + \sum_{\substack{i}}\frac{c_iO_i}{\Lambda^2} .
\label{eqn:eft}
\end{equation}

The sum in Eq. (\ref{eqn:eft}) runs over a total of 59 operators $O_i$. In the absence of a specific high-energy model to match to, the dimensionless Wilson coefficients $c_i$ are all free parameters that must be extracted from data. Clearly, performing a simultaneous fit of 59 free parameters is unfeasible. When one considers a specific class of observables, however, the relevant operator set is much smaller than this. The EFT fit procedure is then straightforward: Firstly, one gathers a set of observables that may be thought of as sensitive to potential new physics effects. Secondly, one writes down all the operators that can affect these observables. Finally, one computes the effects of these operators on the observables and fits to the data, thus extracting values or bounds on their corresponding Wilson coefficients.

The top quark sector is a well-motivated place to look for the effects of dimension-6 operators. The top quark plays a special role in most scenarios that explain electroweak symmetry breaking, scenarios which in turn typically predict modified top couplings which can be modelled by dimension-6 operators, provided the assumptions stated in the first paragraph are satisfied. In addition, the plethora of top quark measurements available from the LHC and Tevatron allows top couplings to be scrutinised with high precision in a global fit. This talk summarises the results of such a fit.

\section{Global fit to Tevatron and LHC Run I data}
With the assumptions of $\mathcal{CP}$-conservation and minimal flavour violation, the complete\footnote{In this analysis, we work at leading order in the EFT; that is, we do not consider operators induced by mixing or loops.} set of operators from Eq. (\ref{eqn:eft}) that modify the couplings of the top quark are

\begin{align}
O^{(1)}_{qq} &= (\bar{q}\gamma_{\mu}q)( \bar{q}\gamma^{\mu}q)  &  O_{uW} &= (\bar{q}\sigma^{\mu \nu} \tau^I u)\tilde \varphi W_{\mu\nu}^{I}  & O^{(3)}_{\varphi q} &= i(\varphi^\dagger \overleftrightarrow{D}^I_\mu \varphi )(\bar{q}\gamma^\mu \tau^I q) \nonumber \\
O^{(3)}_{qq} &= (\bar{q}\gamma_{\mu}\tau^Iq)( \bar{q}\gamma^{\mu}\tau^I q) &  O_{uG} &= (\bar{q}\sigma^{\mu \nu} T^A u)\tilde \varphi G_{\mu\nu}^{A}  & O^{(1)}_{\varphi q} &= i(\varphi^\dagger \overleftrightarrow{D}_\mu \varphi )(\bar{q}\gamma^\mu q)    \nonumber \\
O_{uu} &= (\bar{u}\gamma_{\mu}u)( \bar{u}\gamma^{\mu} u) &  O_{G} &= f_{ABC} G_{\mu}^{A \nu}G_{\nu}^{B \lambda} G_{\lambda}^{C \mu}  &  O_{uB} &= (\bar{q}\sigma^{\mu \nu}u)\tilde \varphi B_{\mu\nu}       \nonumber \\
O^{(8)}_{qu} &=  (\bar{q}\gamma_{\mu}T^Aq)( \bar{u}\gamma^{\mu} T^Au) &  O_{\tilde G} &= f_{ABC} \tilde G_{\mu}^{A \nu}G_{\nu}^{B \lambda} G_{\lambda}^{C \mu}  & O_{\varphi u} &= (\varphi^\dagger i \overleftrightarrow{D}_\mu\varphi)(\bar{u}\gamma^\mu u)    \nonumber \\
O^{(8)}_{qd} &= (\bar{q}\gamma_{\mu}T^Aq)( \bar{d}\gamma^{\mu} T^Ad)  &  O_{\varphi G} &= (\varphi^\dagger \varphi)G_{\mu\nu}^{A}G^{A \mu\nu} \nonumber &  O_{\varphi \tilde G} &= (\varphi^\dagger \varphi) \tilde G_{\mu\nu}^{A}G^{A \mu\nu} \\
O^{(8)}_{ud} &= (\bar{u}\gamma_{\mu}T^Au)( \bar{d}\gamma^{\mu} T^Ad) \,.
\label{eqn:allops}
\end{align}

\begin{figure}[!h]
\includegraphics[width=\textwidth]{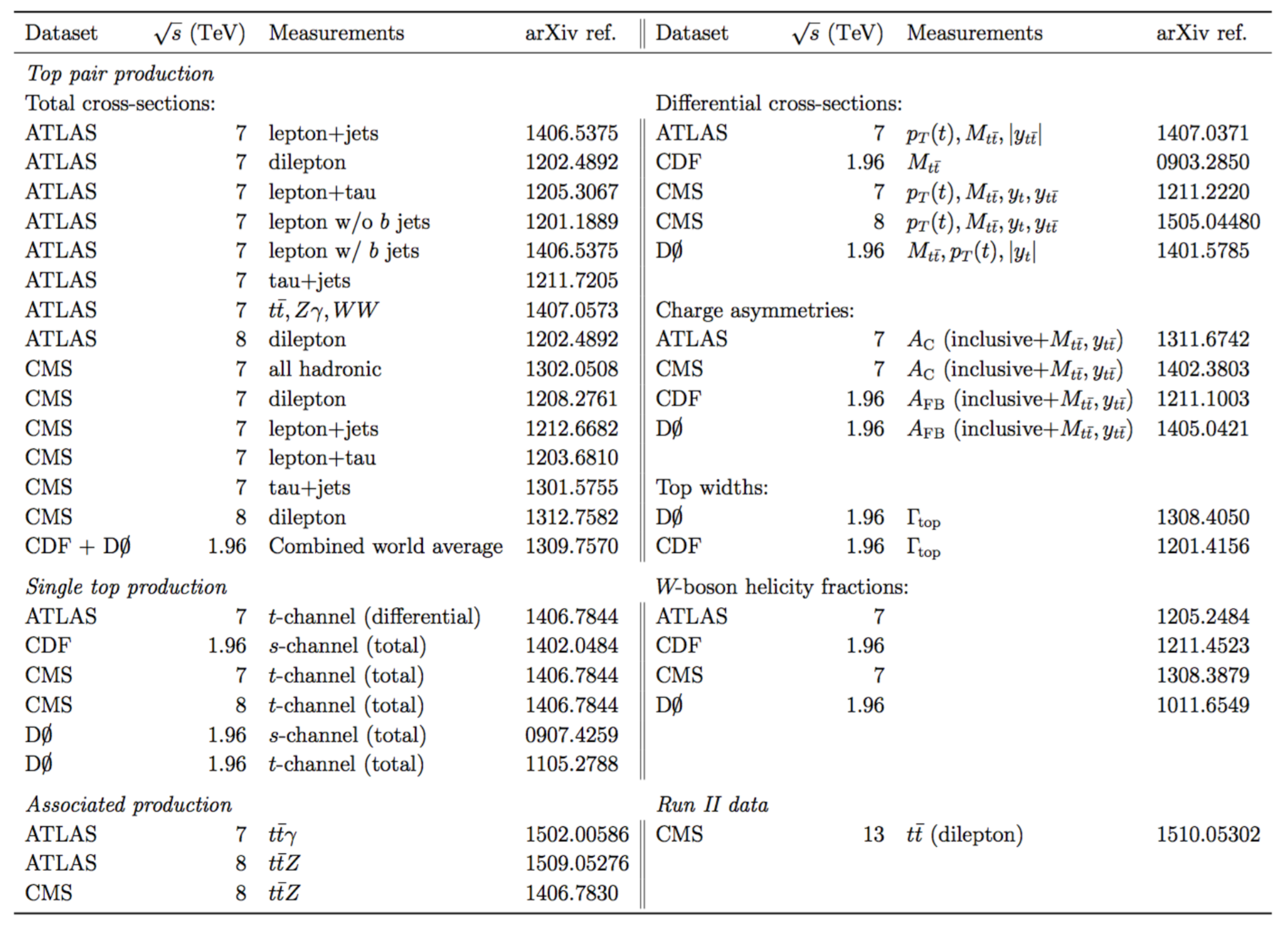}
\caption{The measurements entering our fit. See Ref. \cite{Buckley:2015lku} for more details.}
\label{tab:measurements}
\end{figure}

For a complete discussion of the notation used, and the phenomenology of each operator, the reader is referred to Refs \cite{Buckley:2015nca,Buckley:2015lku}. This operator set of Eq. (\ref{eqn:allops}) was implemented in FeynRules\cite{Alloul:2013bka}, and interfaced via UFO\cite{Degrande:2011ua} to MadGraph\cite{Alwall:2014hca}, which was used to generate theory predictions at each sample point in the parameter space of Wilson coefficients. Higher-order QCD corrections to these observables were taken into account with MCFM\cite{Campbell:2010ff} and MC@NLO\cite{Alwall:2014hca}, with associated scale and PDF uncertainties. Using techniques from Ref. \cite{Buckley:2009bj}, a polynomial function was used to interpolate between each sample point, thus building up a smooth picture of the dependence of each observable used on each operator. A test statistic was constructed; in our case a correlated $\chi^2$, between theory and data, and used to draw exclusion contours on the Wilson coefficients $C_i$. The input measurements used in the fit are summarised in Fig. \ref{tab:measurements}. 

\begin{figure}[!t]
\begin{center}
\includegraphics[width=0.8\textwidth]{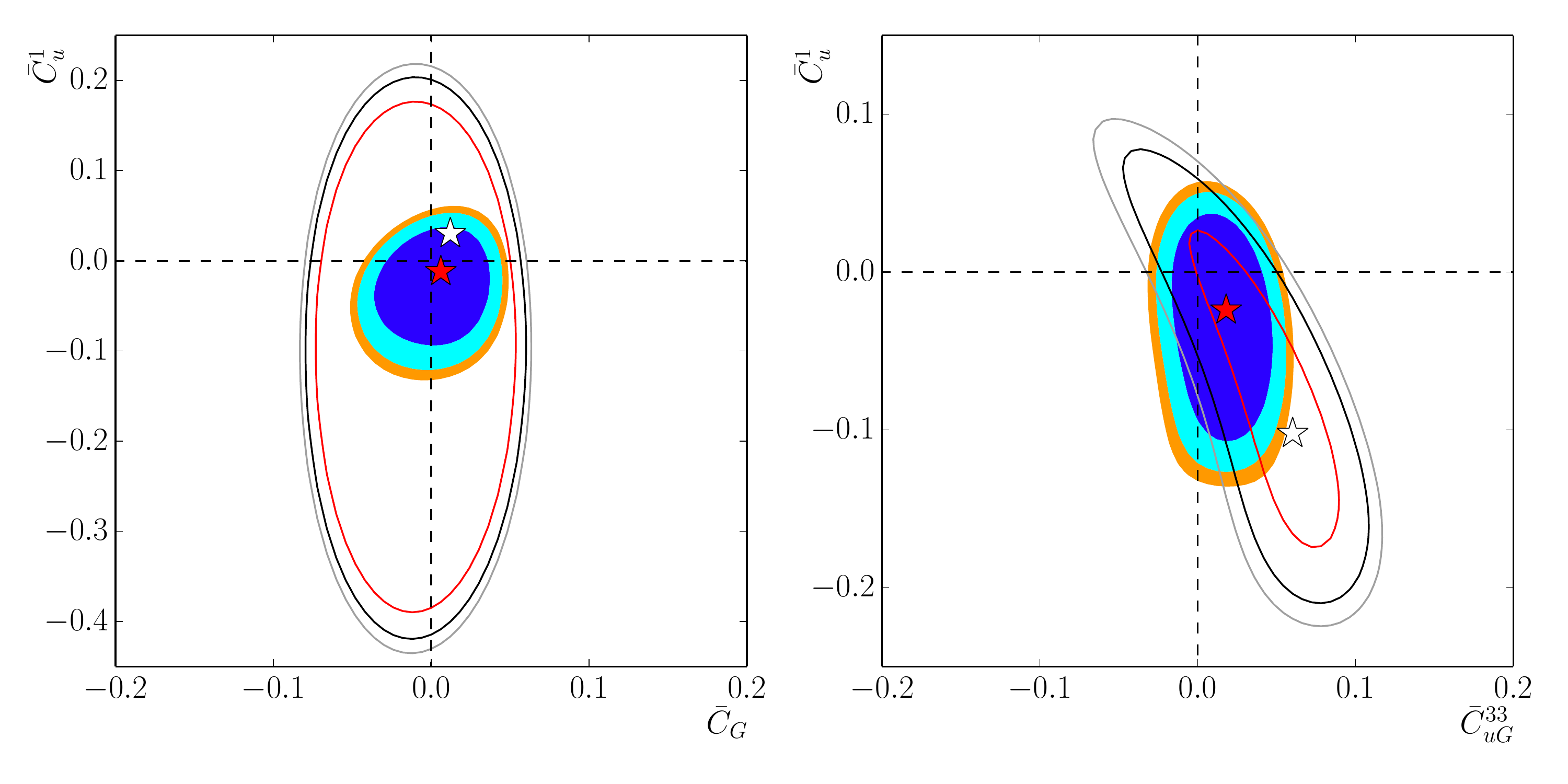}
\caption{2-dimensional 68\%, 95\% and 99\% confidence intervals on pairs of operators contributing to top pair production. Left: lines represent constraints obtained using total cross-sections only; contours are obtained using differential measurements as well. Right: lines are obtained using Tevatron data; contours are obtained using only LHC Run I data.}
\label{fig:contours}
\end{center}
\end{figure}

The main result of our fit is that all operator coefficients are consistent with zero at the 95\% confidence interval. It is interesting to ask which observables have the dominant pull on the fit. We address this question in Fig. \ref{fig:contours}, where we show the joint 68\%, 95\% and 99\% confidence intervals on a pair of operators that modify top pair production. On the left, the lines are the bounds obtained using inclusive cross-section measurements only, whereas the filled contours include differential distributions as well. This shows that differential measurements are the most sensitive observable in constraining $D=6$ operators. On the right, the lines are the bounds obtained using Tevatron $t\bar{t}$ measurements only, whereas the contours are obtained using only LHC Run I data, showing that the LHC has already overtaken the Tevatron in terms of sensitivity, owing to the larger statistical sample collected there.

\begin{figure}[!t]
\begin{center}
\includegraphics[width=0.5\textwidth]{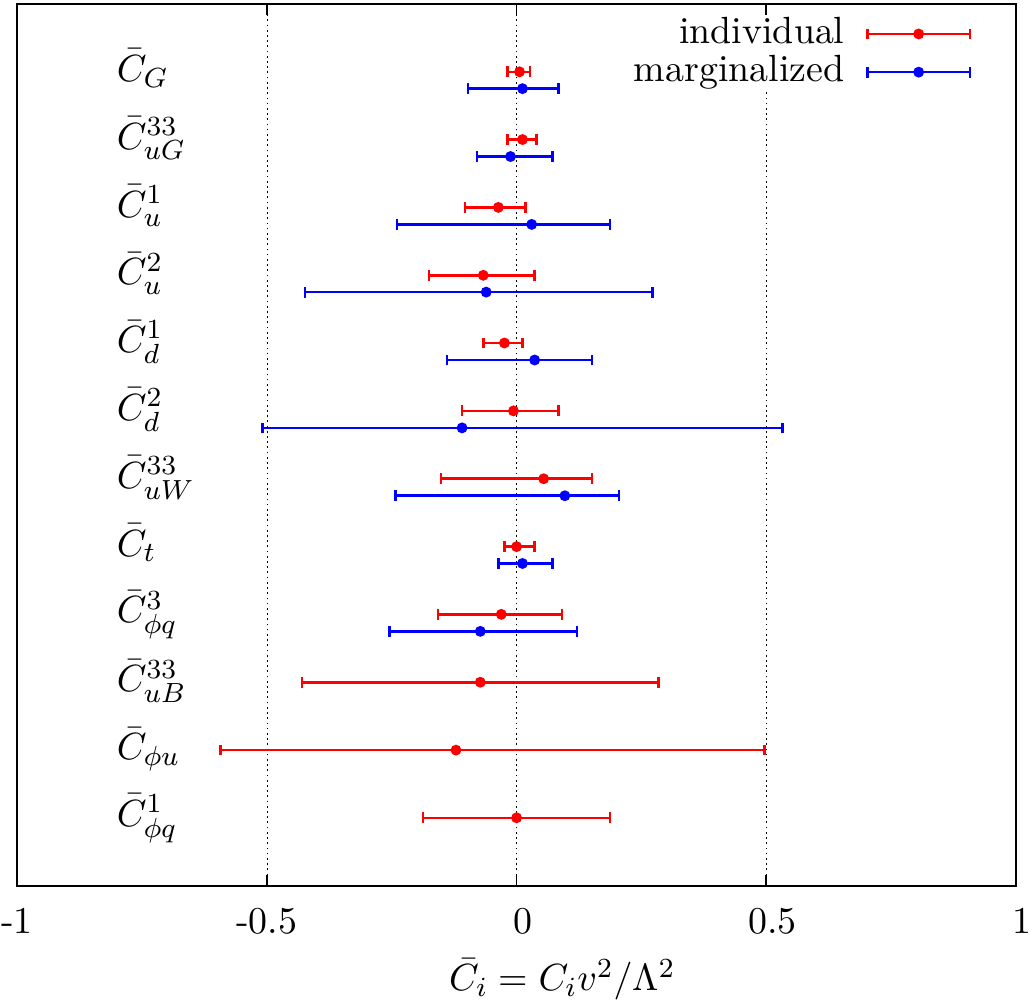}
\caption{Individual (red) and marginal (blue) 95\% confidence limits on the Wilson coefficients for each operator we consider in this fit. Figure taken from Ref. \cite{Buckley:2015lku}.}
\label{fig:constraints}
\end{center}
\end{figure}

In addition to top pair production, we also examine single top production, associated pair production with a $Z$ or a $\gamma$, various charge asymmetries in $t\bar{t}$ production, as well as observables relating to the top quark decay; namely helicity fractions and indirect measurements of the top width. The final 95\% confidence intervals on the operators we consider are presented in Fig. \ref{fig:constraints}. The constraints are presented in two ways: firstly, we fit to each operator one at a time (red), or in a global fit where we marginalise over all the remaining operators (blue). No finite marginal constraint is obtained on the final three operators, due to their much weaker impact on the fit than the rest of the operator set. Again, we refer the reader to Ref. \cite{Buckley:2015lku} for a more detailed discussion of all of these points. 

\section{Improving the fit with boosted observables}
The Wilson coefficient bounds presented in Fig. \ref{fig:constraints} are rather weak, and approach the region where the $D=6$ EFT description may no longer be valid. If we take $C_i = 1$, for instance, we typically find bounds of $\Lambda \gtrsim 500 $ GeV. This scale is resolved by several of the measurements entering our fit; namely the high-energy tails of the kinematic distributions, thus violating the condition $E \ll \Lambda$ that motivated the EFT in the first place.  To try to isolate the regions of phase-space that are most sensitive to $D=6$ operators, we perform an analysis of boosted ($p_T^t > 200$ GeV) top pair production in the semileptonic decay channel, by merging the decay products of the hadronic top into a fat jet, and reconstructing the top $p_T$ using HepTopTagger\cite{Kasieczka:2015jma}. For $p_T^t < 200$ GeV we perform a standard resolved analysis. We fit the operators of Eq. (\ref{eqn:allops}) that contribute to $t\bar{t}$ production to Standard Model pseudodata in both the resolved and boosted regions. Theory uncertainties are treated as before. On the experimental side, we take statistical uncertainties corresponding to 30 fb$^{-1}$ of SM data, and assume a 20\% statistical uncertainty on each bin. We then ask: what improvements can be made on current bounds when systematic and statistical uncertainties are reduced?

\begin{figure}[!t]
\begin{center}
\includegraphics[width=0.8\textwidth]{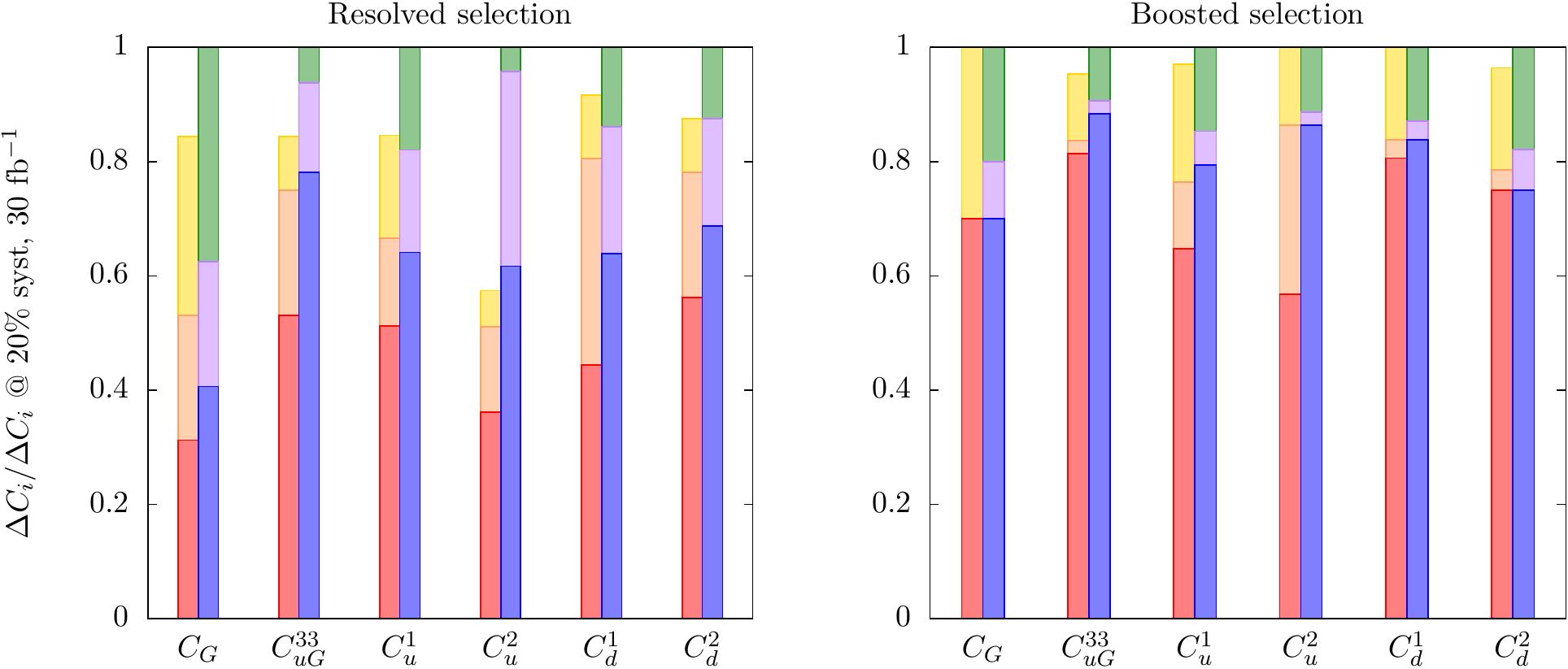}
\caption{Fractional improvement on the 95\% confidence intervals for the $t\bar{t}$ operators. The green bar represents the 95\% confidence limit obtained from 20\% systematic uncertainty and 30 fb$^{-1}$ of data. The purple and blue bars represent the fractional improvement with respectively, 300 fb$^{-1}$ and 3 ab$^{-1}$ of data, also at 20\% systematics, while the yellow, orange and red are the analogous data sample sizes for 10\% systematics. Figure taken from Ref. \cite{Englert:2016aei}.}
\label{fig:norm_constraints}
\end{center}
\end{figure}

The results of this analysis are shown in Fig. \ref{fig:norm_constraints}. The conclusions are quite different for the resolved and the boosted selections. For the former, improvements can be made both by taking more data and by reducing systematics, but reducing systematics is more important. Taking $C^1_u$ as an example, we see that improving systematics by 10\% and taking 300 fb$^{-1}$ of data produces a comparable bound to keeping current systematics and taking a tenfold larger data sample. For the boosted case, on the other hand, we see that, at low statistics there is no gain from reducing systematics, which merely reflects that high-$p_T$ boosted measurements are statistics limited at this stage. Even with larger data samples, the gain to be made by reducing systematics is more modest. This says that another input to the fit becomes important: the theoretical modelling of the high-$p_T$ tail. Approaching 3000 fb$^{-1}$, theory uncertainties will become the main driving force in improving the top EFT fit at the LHC. Therefore, there is work to be done on both sides if the LHC is to reach its full potential for measuring or placing limits on $D=6$ operators.

\section*{Acknowledgements}
The speaker wishes to thank the organisers of the CKM workshop, especially the convenors of the top working group; Veronique Boisvert and Joachim Brod, for the pleasant and productive atmosphere throughout.

\bibliographystyle{JHEP}
\bibliography{bibfile}

\end{document}